\documentclass[cits]{PoS}
\newcommand{\dd}{\mbox{d}}
\def\spose#1{\hbox to 0pt{#1\hss}}

\def\lta{\mathrel{\spose{\lower 3pt\hbox{$\mathchar"218$}}
\raise 2.0pt\hbox{$\mathchar"13C$}}}
\def\gta{\mathrel{\spose{\lower 3pt\hbox{$\mathchar"218$}}
\raise 2.0pt\hbox{$\mathchar"13E$}}}

\def\setR{\mathbb{R}}
\def\setC{\mathbb{C}}

\newcommand{\Ka}{{\cal K}}
\newcommand{\GN}{G_{_\mathrm{N}}}
\newcommand{\mP}{M_{_\mathrm{Pl}}}

\newcommand{\lP}{\ell_{_\mathrm{Pl}}}
\title{Observing alternatives to inflation}

\ShortTitle{Bouncing cosmology}

\author{\speaker{Patrick Peter}\\
  ${\cal G}\setR\varepsilon\setC{\cal O}$ -- Institut
  d'Astrophysique de Paris, UMR7095 CNRS \\
  Universit\'e Pierre \& Marie Curie \\ 98 bis boulevard Arago \\ 75014 Paris, France\\
  E-mail: \email{peter@iap.fr}}

\abstract{We discuss the possibility that the inflationary paradigm,
  undoubtfully today's best framework to understand all the present
  cosmological data, may still have some viable challengers. The
  underlying idea for such discussions is that although inflation
  already passed quite a large number of tests, indeed enough to make
  it part of the so-called ``standard model'' of cosmology, it has
  always been through indirect measurements: there is not a chance
  that we may ever directly check its validity, and therefore, in
  order to assert its factuality with increasing level of confidence,
  it is required that we compare its predictions not only to
  observations, but also to as many contenders as possible. Among
  other categories of possible models, we wish to put the emphasis in
  particular on bouncing cosmologies that, however not as complete as
  the inflation paradigm might be, could still represent a reasonnable
  way of explaining the current data. Hopefully, future data will be
  able to discriminate between these various sets of theories.}

\FullConference{International Workshop on Cosmic Structure and Evolution - Cosmology2009,\\
		September 23-25, 2009\\
		Bielefeld, Germany}

\begin{document}

\section*{Introduction}

Standard, pre-1980, cosmology, was plagued with many problems. Among
those, most have received solutions in the framework of what
eventually became known as the inflation paradigm
\cite{Linde-book,inf25,PPJPU}. Those problems were that of the
existence of a singularity (not completely solved
\cite{Borde-Vilenkin-PRL}), the question of why the properties of the
cosmic microwave background (CMB) were the same over distances much
larger than the horizon, and the fact that the Universe is observed to
be flat \cite{Guth} and homogeneous (this problem, although
considerably alleviated by a phase of inflation, can be argued not to
be actually solved by this phase) \cite{homogeneity,inhom}. These are
unavoidable difficulties coming directly from General Relativity (GR)
and the Friedmann-Lema\^{\i}tre-Robertson-Walker (FLRW) metric
whenever one assumes the Universe to be filled with a perfect fluid
(dust or radiation in practice).

A different kind of problem is that of monopoles \cite{Preskill79},
namely particle-like topological defects \cite{VS00} (TD) with large
mass and magnetic interactions whose density in the early Universe can
exceed by far the flatness limit, implying a very rapid crunch right
after the big-bang, thus in disagreement with the very basic
observation that the Universe merely exists! Also seemingly
unavoidable, this problem relies heavily on the existence of a grand
unification \cite{GUT} for the interactions, unification which,
although very seriously expected on theoretical grounds, is by no
means as certain as GR and its FLRW solution.  While solving these
delicate matters, including the monopole excess problem, inflation
provided, as an ``unexpected'' bonus, a means of calculating the
expected spectrum of primordial perturbations \cite{MC812,MFB},
providing a reasonable theory for the initial conditions of
primordial perturbations.

Finally, there still is a set of cosmological puzzles which are not
addressed within the inflationary paradigm. Those are the existence of
dark matter \cite{DM}, and the observation that the Universe is
currently accelerating, which is interpreted as a yet-unknown
constituent, called dark energy \cite{DE}, assumed to be either a mere
cosmological constant or a more elaborate quintessence field. Lastly,
there is the very large asymmetry between matter and anti-matter that
fits nowhere in the standard model of particle physics; this, however,
may turn out to be more a problem of this latter field rather than of
cosmology \cite{baryo}. This last category is not solved by inflation,
although there are models aiming at addressing those in unifying
frameworks.

So the current situation is the following: inflation, as a paradigm,
solves most of the cosmological puzzles in a consistent way, using GR
and scalar fields (semi-)classically, and it can be implemented in high
energy particle physics theories \cite{HEP_inf}. From the
observational standpoint, it is fair to say that is also make
falsifiable predictions (the spectrum for the density perturbations
and, as a consequence, the CMB temperature anisotropy distribution),
all these having been shown to be consistent with the data. It turned
out, more recently, that is was also possible to implement an
inflationary phase in a string \cite{stringTH} framework
\cite{stringflation}. Inflation thus appears like a cosmological
panacea, and one wonders why one would even consider alternatives.

The first reason to discuss noninflationary scenarios is quite simple:
although inflation can be implemented in various theories, it is not
the only possible cosmological outcome of these theories. Other
scenarios turn out to be possible, not all of them being excluded by
the data. Thus, there are challengers worth investigating. Besides,
working out a challenger permits comparisons and different
predictions. Future observations will then discriminate between
models, as past observations did: would it be merely to enforce the
inflation paradigm, challengers would be already very useful.

Yet another reason to be concerned with noninflationary cosmology is
the fact that it has problems of its own \cite{RB25}. First of all, it
does not address the question of the primordial singularity, although
it is not clear that it is a question. More to the point are technical
problems that need to be fixed. For instance, most models of inflation
is heavily based on a fundamental scalar field ... yet no such field
has ever been observed! This is a problem shared with many other
theories, as in particular in particle physics the only missing bit
is, precisely, the scalar Higgs particle. Some might also argue that
using GR at energies that close to the Planck/string scale (the
typical inflation scale is $E_\mathrm{inf} \gta 10^{-5} \mP$, while
the value of the scalar field responsible for the occurrence of
inflation can even be larger than the Planck scale \cite{Lyth}) is
questionable. Then comes a sort of hierarchy problem, by which one
means the fact that at least one dimensionless number in the scalar
field potential should be given a fine-tuned small value, of the order
of $10^{-12}$; again, whether this is a negative point is arguable but
should be kept in mind. Finally, one sets initial conditions at a time
when the wavelength $\lambda = k^{-1}/a(t)$ [$a(t)$ being the scale
factor depending on time $t$, for a given comoving wavenumber $k$] of
the perturbations can be smaller than the Planck length $\lP$ itself;
this is disturbing as one could imagine unknown and uncontrollable
quantum gravity effects should be important, if not dominant, at such
scales \cite{TransP}.

Having established the need for alternatives, the question remains as
to which alternative? Old models based on TD (cosmic strings in
particular) were found to be plausible, but only to partly
\cite{CSCMB} explain the large scale structure, i.e. at the 10~\%
level (and then, some argue, with a slightly better fit). In that
sense, TD cannot really be seen as an alternative.

Actual alternatives that claim to be compatible with the current data
are not so many. The first, historically, called the Pre-big bang
(PBB) \cite{PBB}, was based on string theory, using its dualities; the
PBB is hard to reconcile with the data unless curvaton-like mechanism
is taking place. Yet another mechanism, also involving string theory,
was the so-called ekpyrotic scenario \cite{ekp}, which is subject to
quite some controversy \cite{noekp}. All these scenarios share the
common feature of having a bouncing phase, at least in the Einstein
frame. Therefore, it seems reasonable, as a generic alternative to
inflationary cosmology, to consider bouncing cosmology, whose history
is in fact much older than that of inflation, tracing back to the
1930's \cite{historyBounce}! We shall not describes, in the short
review below, models such as those based on string gas cosmology: they
are as worth investigating as those presented below, but apart from
the fact that space is lacking, I have never studied them in
sufficient details to pretend discussing them here; I therefore refer
the interested reader to Ref.~\cite{RB25} and the references given
there for details (as well as another view on bounces and inflation
caveats).

\section{Standard failures and some solutions}

The standard cosmological model is based on the FLRW metric \cite{PPJPU}
\begin{equation}
\dd s^2 = -\dd t^2 + a^2(t) \gamma_{ij} \dd x^i \dd x^j = a^2(\eta)
\left( -\dd\eta^2 + \gamma_{ij} \dd x^i \dd x^j\right),
\label{FLRW}
\end{equation}
where $\gamma_{ij}=\left( 1 +\frac14 \Ka \vec{x}^2\right)^{-1}
\delta_{ij}$ is the spatial metric; Eq.~(\ref{FLRW}) defines the
conformal time $\eta$ as a function of the cosmic time $t$ and the
scale factor $a$ through $\dd t = a \dd\eta$. Plugging this form of
the metric into Einstein equations with the stress-energy tensor for a
perfect fluid with equation of state $w$, namely
\begin{equation}
T_{\mu\nu} = \left( \rho + p \right) u_\mu u_\nu + p g_{\mu\nu},
\label{Tw}
\end{equation}
with energy density $\rho$, pressure $p=w\rho$ and $u_\mu$ is the
fluid 4-velocity satisfying $u_\mu u^\mu=-1$, one finds that since
$\nabla_\mu T^{\mu\nu}=0$ implies $\rho \propto a^{-3 (1+w)}$, the
solution for the scale factor reads (for $\Ka=0$)
\begin{equation}
a \propto t^{2/[3(1+w)]} \propto \eta^{2/(1+3w)}, 
\label{singularity}
\end{equation}
provided $w\not=-1$. For a dust-dominated Universe with $w=0$, this is
$a_\mathrm{dust}\sim t^{2/3} \sim \eta^2$, while the
radiation-dominated case $w=\frac13$ is $a_\mathrm{rad}\sim t^{1/2}
\sim \eta$: in both cases, the limit when $t,\eta\to0$ is singular,
namely $a\to0$. That is the origin of the singularity problem. In a
bouncing cosmology, one assumes that, for whatever reason to be
determined afterwards, the scale factor is bounded from below, so
that, at all times, $a\not=0$. As a result, the singularity problem is
merely a non issue in the bouncing case ... but then of course, the
question remains of the matter content which permits such a scale
factor behavior.

The horizon problem is slightly more involved, although not that much
\cite{PPNPN3}. To recall, the particle horizon at a point $P$ (the
observer, say) is that surface dividing all points in the Universe
into two distinct families, namely those that have already been
observed at time $t$ by $P$, and those that have not. Specifically,
integrating along the past light-cone at the point $P$, all the way to
the origins of time $t_\mathrm{i}$, one finds a particle horizon
exists provided the integral
\begin{equation}
d_{_\mathrm{H}} = a (t) \int_{t_\mathrm{i}}^t \frac{\dd\tau}{a(\tau)}
\label{horizon}
\end{equation}
converges. The quantity $d_{_\mathrm{H}}$ is called the horizon, and the
problem in standard cosmology is that not only is it finite (the
integral converges for a perfect fluid as before provided
$w>-\frac13$), but also, in comparison to the Hubble scale, it is
small. In the framework of a bouncing Universe, one assumes that there
is no such thing as the origin of time, and hence one sends
$t_\mathrm{i}$ to negative infinity. If, during the contracting phase,
one also have domination by a perfect fluid with equation of state
$w>-\frac13$, then the integral diverges, and so it remains infinite
for all subsequent times, whatever happens at the bounce: all points
in the Universe have, at some stage, been in causal contact.

The flatness problem now mostly stems from the equation which expresses
the density $\rho(t)$ relative to the critical density
$\rho_\mathrm{c}$. With $H\equiv\dot{a}/a$
(a dot meaning $\dd /\dd t$), this is
\begin{equation}
\frac{\dd}{\dd t} |\Omega -1|=-2\frac{\ddot{a}}{\dot{a}^3}, \ \ \ \ \
\ \ \ \Omega \equiv \frac{\rho(t)}{\rho_\mathrm{c}(t)} = \frac{8\pi\GN
  }{3 H^2(t)} \rho(t).
\label{crit}
\end{equation}
For a nonaccelerating ($\ddot{a}\geq0$) expansion ($\dot{a}>0$), it is
clear that $\Omega$ is always moving away from unity: the flat
Universe ($\Omega=1$) is unstable. Inflation solves this problem by
having a phase of accelerated expansion ... the bouncing paradigm
suggests just the opposite, namely that $|\Omega-1|$ decreases because
of a nonaccelerated contraction that lasted sufficiently long to
compensate for the currently ongoing expansion.

With homogeneity, the situation is more intricate. Inflation proposes
a dynamical mechanism by which any pre-existing inhomogeneity and
anisotropy is washed out: an extremely small homogeneous region is
almost instantaneously expanded into a size much larger than the
current Hubble scale. Although this tremendously alleviates the
problem, it remains to assume that there was a sufficiently
homogeneous region of the required size to begin with
\cite{homogeneity} (see however Ref.~\cite{inhom} for a
counter-argument). Whether this purely classical requirement is
actually necessary may be still debatable, it is however not
immediately obvious what this implies at the quantum level since
homogeneous regions presumably do have a non-negligible probability to
occur in the first place in a (yet-unknown) quantum gravity setup. For
the bouncing situation, more information on the initial condition is
needed.

At the origin of the contracting phase, the Universe is supposed to be
large and extremely dilute, so both the stress-energy and the Ricci
tensor are, by virtue of Einstein equations, very small. Adding the
requirement that the Weyl tensor should also be small (i.e. we assume
the Weyl curvature hypothesis \cite{weyl}), we ensure the geometry to
be almost flat to begin with, and look at the growth of whatever
initial inhomogeneities we have have started with.

Since we assume a rarefied Universe, any inhomogeneity, even a large
one having $\delta\rho/\rho\gtrsim 1$, has negligible self-gravitation
because $\rho$ is small. In much the same way as it is true that sound
waves do not condense, these initial inhomogeneities must
dissipate. Let us now restrict attention to the (relevant)
dust-dominated contraction for which the Jeans length is
$\lambda_\mathrm{J} = c_\mathrm{s} \sqrt{\pi/(\GN\rho)}\propto
a^{3/2}$, with $c_\mathrm{s}$ the sound velocity, which in this
case is essentially the equation of state parameter $c_\mathrm{s} \sim
w\ll 1$: we cannot at this level approximate $w\simeq 0$ as this would
imply a vanishing Jeans length and hence exponential growth of all
scales! in fact, provided $w\not=0$, which it cannot strictly be since
there will always, in practice, be some amount, however tiny, of
interaction between particles, it is always to set the initial
sufficiently backwards in time to have the scale factor, hence the
Jeans length, as large as one wants. We thus assume
$\lambda_\mathrm{J}$ to be larger than any scale than any scale of
cosmological relevance today.

The dust velocity evolving as $v\propto a^{-1}$ and the number density
as $n\propto a^{-3}$, the mean free path is $\lambda_{_\mathrm{MFP}} =
(n\sigma)^{-1} \propto a^3$, with $\sigma$ the interaction
cross-section between dust particles. Again, although $\sigma$ is
expected to be small for the dust approximation to be valid, it cannot
vanish entirely. For a given wavelength $\lambda$, the associated
dissipation time is
\begin{equation}
t_\mathrm{d} (\lambda) = \frac{\lambda}{v} \left( 1+
  \frac{\lambda}{\lambda_{_\mathrm{MFP}}} \right),
\label{td}
\end{equation}
which should be compared with the Hubble time $t_{_\mathrm{H}}$.

The dust scale factor relates with the Hubble scale $R_{_\mathrm{H}}$
through $a\propto R_{_\mathrm{H}}^{2/3}$, and hence
$\lambda_{_\mathrm{MFP}} = R_{_\mathrm{H}}^2/L$, with $L$ an unknown
constant with the dimension of a length, in principle to be determined
by the microphysics at the time at which we want to fix the initial
conditions. Taking the wavelength to be a fixed fraction $x$ of the
Hubble scale now $R_0$, i.e. $\lambda = x R_0$, and similarly the
Hubble radius at the time considered to be $R_{_\mathrm{H}} = yR_0$,
one finds that the ratio of the dissipation to the Hubble times is
\begin{equation}
\frac{t_\mathrm{d} (\lambda)}{R_{_\mathrm{H}}} \propto x y^{-1/3}
\left( 1 + \frac {L}{R_0} \frac{x}{y^2}\right), 
\label{Hdis}
\end{equation}
so that it suffices to take initial conditions sufficiently far in the
past when the Universe was much larger than it is now, i.e., to take
$y\gg 1$ (with $x$ of order unity and $L$ of order the present day
Hubble scale), to find $t_\mathrm{d}\ll R_{_\mathrm{H}}$: in a
characteristic Hubble time $R_{_\mathrm{H}}$, the system has had
plenty of characteristic dissipation times, and so has dissipated
completely before gravity begun to play any role. Actually, provided
the Universe spends a sufficiently long time in this contracting
phase, the diffusion is so effective that only quantum fluctuations
due to the uncertainty principle could survive: vacuum initial
conditions for the perturbations is a consequence of the
homogenisation mechanism.

\section{Modelling the bounce}

GR with ordinary fluids does not admit bounce solutions. Indeed,
the metric (\ref{FLRW}) with the fluid (\ref{Tw}) provide the
Einstein (Friedmann) equations (in the absence of cosmological constant)
\begin{equation}
  \left(\frac{a'}{a}\right)^2 + \Ka = \frac{8\pi\GN}{3} a^2 \rho \ \ \ \
  \ \ \hbox{and} \ \ \ 
  \ \ \ \frac{a''}{a} = \frac{8\pi\GN}{6} a^2 \left( \rho -3 P\right) - \Ka,
\label{Fried}
\end{equation}
where ${}'\equiv\dd/\dd\eta$. A bounce being
defined as the point at which the scale factor reaches a minimum
value, it means one must have, at the bounce conformal time,
$a'_\mathrm{b} = 0$ and $a''_\mathrm{b} > 0$. Unless $\Ka >0$, the
first condition requires that the energy density for the matter
vanishes at that time. It is therefore not surprising that most
classical bouncing models relies on closed spatial sections
\cite{modK}. Other kinds of models either contain negative energies
\cite{negE} (and can therefore only by analyzed as effective) or
contain nonstandard terms \cite{K}. Although all of these models have
their specific interests, I would like to concentrate on yet another
category, based on quantum cosmology \cite{QC}.

The model thanks to which one manages to perform a bounce and produce
a scale invariant spectrum of scalar perturbations (see the next
section) turns out to be the simplest possible one, namely one for
which one gravity plus a single perfect fluid with almost vanishing
equation of state. When quantum cosmological effects are taken into
account, it is no longer necessary to demand positive spatial
curvature, and therefore we can set $\Ka=0$.

In quantum cosmology, separating the background and the perturbations
means factorizing the Universe wavefunction $\psi$ into zeroth and
second orders as
\begin{equation}
\psi = \psi^{(0)}\left(a,\eta \right) \times \psi^{(2)}\left[ a,\Psi
  \left(\vec{x}\right),\Phi \left(\vec{x}\right), h_{ij}\left(\vec{x}
  \right),\eta\right].
\label{wave}
\end{equation}
The idea is to solve the Wheeler-de Witt equation for the background
wavefunction first, and using a Bohmian interpretation \cite{Bohm},
derive the quantum trajectory for the scale factor. One finds, for the
scale factor
\begin{equation}
a\left(\tau\right) = a_0 \left[
  1+\left(\frac{\tau}{\tau_0}\right)^2\right]^{1/[3(1-w)]} \ \ \ \ \ \
  \hbox{with} \ \ \ \ \  \dd\eta =
  \left[a\left(\tau\right)\right]^{3w-1}\dd\tau, 
\label{atau}
\end{equation}
where $a_0$ and $\tau_0$ are two arbitrary integration constants. They
represent respectively the minimum value of the scale factor and the
bounce duration. In terms of such a solution, one now needs to
consider metric perturbations. This we do in the next section.

\section{Spectrum of perturbations}

In section 1, we showed that the initial conditions for the
primordial fluctuations should be vacuum like. With this in mind, we
shall now propagate these perturbations through the bounce itself in
order to determine the observational predictions of the bouncing
paradigm.

The metric, one step beyond the background level described above, now
reads 
\begin{equation}
\dd s^2 = a^2(\eta)\left\{-\left(1+2\Phi\right) \dd\eta^2
  +\left[\gamma_{ij} \left(1-2\Psi\right) + h_{ij} \right] \dd x^i \dd
x^j\right\},
\label{gpert}
\end{equation}
in the longitudinal gauge (we do not consider here the possible vector
part). The tensor $h_{ij}=h_{ji}$ is divergenceless ($\nabla_i
h^{ij}=0$) and traceless ($\gamma^{ij}h_{ij}=0$); spatial indices are
raised and lowered with the background spatial metric
$\gamma_{ij}$. As for the perturbed stress-energy tensor components,
they are
\begin{equation}
\delta T_{00} = \rho a^2 \left(\frac{\delta\rho}{\rho} +2\Phi\right),
\ \ \ \delta T_{0i} = -\rho a^2 (1+w) \nabla_i v , \ \ \ \ \delta T_{ij} = P a^2 \left( h_{ij} -2
    \gamma_{ij} \Psi + \frac{\delta p}{p}\gamma_{ij} +\pi_{ij} \right),
\label{Tpert}
\end{equation}
where the anisotropic stress tensor $\pi_{ij}$ can be further
decomposed into scalar and tensor part through
$\pi_{ij}=\left(\nabla_i\nabla_j-\frac13 \gamma_{ij}\Delta\right)
\overline{\pi}+\overline{\pi}_{ij}$, the latter part
$\overline{\pi}_{ij}$ being divergenceless and traceless.

We have not considered vector modes: those, as is well known, are not
really dynamical and, in an expanding Universe, are rapidly washed
out if no anisotropic stress is present to begin with. In a
contracting Universe, the situation is far from being that clear, as
one would for instance expect those to grow (in practice with the
square of the inverse scale factor), thus potentially breaking the
perturbation approximation, thereby destroying the whole picture as a
viable cosmological model \cite{vecbounce}. However, such a dramatic
effect depends not only on the specific details of the contracting and
bouncing phases, but also on the initial conditions which, as
discussed above, may simply imply the vector modes to identically
vanish. This is what we assume in what follows. 

Let us now come back to the quantum cosmology situation depicted in
the previous section and plug the solution (\ref{atau}) into the
equation for the second order part of the wavefunction (note that in
this very simple case, one also has that the two potentials appearing
in the metric expansion reduce to a single one, namely
$\Phi=\Psi$). Fourier expanding in terms of the comoving wavenumber
$k$, one finds the usual mode equation for the Mukhanov-Sasaki
variable $v$, related to the Bardeen potential $\Phi$ through
\begin{equation}
\Delta\Phi = -\frac32 a\lP^2\sqrt{\frac{\rho+p}{w}}\frac{\dd}{\dd\eta}
\left( \frac{v}{a}\right),
\label{MS}
\end{equation}
namely
\begin{equation}
v_k'' + \left( c_\mathrm{s}^2 k^2 -\frac{a''}{a} \right) v_k = 0,
\label{mode}
\end{equation}
where the sound velocity $c_\mathrm{s} = \sqrt{w}\ll1$ cannot
vanish. During the contracting phase, the ``potential'' $V=a''/a$ of
this Schr\"odinger-like equation is important close to the bounce: for
sufficiently large negative $\eta=\eta_\mathrm{ini}$, $V\ll
c_\mathrm{s}^2 k^2$, and one can impose vacuum initial conditions,
namely set
\begin{equation}
v_k \left(\eta < \eta_\mathrm{ini}\right) = \frac{\mathrm{e}^{-i
      c_\mathrm{s}k\eta}}{\sqrt{2 c_\mathrm{s} k}},
\label{vkini}
\end{equation}
and subsequently evolve this function with Eq.~(\ref{mode}). The
scalar spectrum $\mathcal{P}_\Phi \propto k^3 |\Phi_k|^2 \propto
A_{_\mathrm{S}} ^2 k^{n_{_\mathrm{S}}-1}$ is then obtained once
$\Phi$, reconstructed from $v$, reaches, in the expanding phase, a
constant value: this is the dominant mode.

The spectrum for the gravitational wave (tensor mode) is obtained in a
completely similar way. In fact, expanding $h_{ij}$ on a basis on
time-independent polarisations modulated by an amplitude $\mu = h/a$,
and going to Fourier space, one finds
\begin{equation}
\mu_k'' + \left( k^2 -\frac{a''}{a} \right) \mu_k = 0,
\label{mu}
\end{equation}
which, apart from the unimportant factor of the sound speed, is the
same as Eq.~(\ref{mode}). The initial condition can be set in an
exactly similar way, so as the dynamics is also the same, one expects
that the tensor spectrum $\mathcal{P}_h \propto k^3 |h_k|^2 \propto
A_{_\mathrm{T}} ^2 k^{n_{_\mathrm{T}}}$  will be, up to a
normalisation factor, equal to the scalar spectrum.

The actual solution which we found, either by numerically integrating
the mode equation or by making approximations and matchings, is that
the spectral indices satisfy
\begin{equation}
n_{_\mathrm{T}} = n_{_\mathrm{S}} - 1 = \frac{12 w}{1+3w} \ll 1.
\label{n}
\end{equation}
Normalizing to the actual data, i.e. setting $A_{_\mathrm{T}} ^2 =
2.08\times 10^{-10}$, one can then calculate the unknown parameters
$a_0$ and $\tau_0$. Those provide the curvature at the bounce itself,
and we find
\begin{equation}
\tau_0 a_0^{3w} \sim 10^3 \lP,
\label{curv0}
\end{equation}
which turns out to be precisely the length scale at which one expects
quantum gravity corrections to become important, while the full
quantum theory is still unnecessary: the Wheeler-de Witt equation is
supposed to make sense in this regime, and our result is robust.

In this category of models, one also finds that the tensor-to-scalar
ratio is related to the scalar spectral index, although in a different
way from the usual inflation case: we find $(T/S) \simeq 4\times
10^{-2} \sqrt{n_{_\mathrm{S}}-1}$. This result is to be compared with
the so-called consistency relation in single field inflation, which
essentially states that the ratio $(T/S)$ ought to be proportional to
$n_{_\mathrm{S}}-1$ and not its square root. This, together with the
fact that $n_{_\mathrm{S}}>1$, again in contradiction with the
inflation prediction,  provides another way of testing either
inflation or one of its contenders.

This kind of models is, of course, not without drawbacks. In
particular, the minimum value of the scale factor, for instance, is
extremely large, $a_0 \gta 10^{20}\lP$, which implies a conceptual
problem as a number of this order of magnitude is rather unexpected,
and in fact the probability, calculated from the wavefunction, for the
occurrence of such a large value, is more than tiny (of the order of
$\mathrm{e}^{-10^{89}}$ in extreme cases!). Generalized solution for
the background wavefunction, describing a moving Gaussian with
nonvanishing velocity term along the $a$-axis, provide a means of
solving such problems \cite{NPN}.

\section*{Conclusions}

I have presented a very rapid tour of the reasons why it is still
possible to find consistent cosmological models that do not include a
phase of inflation, insisting in particular on the simplest
possibility, sticking to 4 dimensional GR-based (and its quantum
version) theories. Other, more involved, models are possible, and many
share with the above most of their characteristic features. Others
still exhibit different properties such as oscillations in the
predicted spectrum of primordial perturbations. Increased precision
experiments such as Planck, whose first scientific data are expected
very soon, will be hopefully able to discriminate between the various
theories discussed here and the more standard inflation paradigm.

Concentrating on a very specific model based on the Wheeler-de Witt
equation, using a Bohmian interpretation for the quantum, I have
summarized the results obtained through the years that tend to show
that plain GR plus a perfect field can produce a sound model of the
early Universe. Of course, this model is most definitely
oversimplified. However, it has the advantage to be able to reproduce
most of the data without introducing arbitrary scalar fields. Although
clearly not perfect and far too simple as yet, it exemplifies the fact
that alternatives are still viable and deserve further investigations.

Further work is necessary in order to assert whether such models can
be made more than viable, but actually competitive alternatives. Like
inflation models, most bouncing scenarios have been developed within
the framework of a single fluid/field component; as the Universe
contains more than one such component, the very first generalisation
one needs to implement is that in which both matter and radiation (and
even perhaps a cosmological constant) are present at the same time. In
brief, one should try and build a scenario that really accounts for
all observations. And as far as future data are concerned, for
instance, it is of paramount interest to understand what is the amount
of nongaussianity they tend to produce, and in what
configurations. Again, this will have to be compared with generic
inflationary predictions, and then with observations, whenever
available.

\section*{Acknowledgments}

I wish to thank the organizers of the meeting that turned out to be
extremely fruitful and interesting. Moreover, I want to warmly thank
all my collaborators on the topics I reviewed here, namely R.~Abramo,
J.~Fabris, F.~Falciano, M.~Lilley, J.~Martin, E.~Pinho, N.~Pinto-Neto,
D.~Schwarz and I.~Yasuda. I wish to thank especially J.~Martin for his
careful reading of the manuscript.

\end{document}